\documentstyle[sprocl]{article}

\def\Journal#1#2#3#4{{#1} {\bf #2}, #3 (#4)}

\def\CJP{\em Can. J. Phys.}

\def\PLB{{\em Phys. Lett.}  B}

\def\be{\begin{equation}}
\def\ee{\end{equation}}
\def\bea{\begin{eqnarray}}
\def\eea{\end{eqnarray}}
\def\vdenom#1#2{{\left(\frac{(#1_1-#2_1)^2}{1-v^2}+ (#1_2-#2_2)^2+
(#1_3-#1_3)^2\right)}}

\begin{document}

\title{CONSTRUCTION AND CONSEQUENCES OF COLOURED CHARGES IN 
QCD\footnote{Published in the proceedings of the AUP workshop {\sl 
Quantum Chromodynamics: Collisions, Confinement and Chaos}, Ed.'s H.M. 
Fried and B. M\"uller (World-Scientific 1997).}}

\author{ E.\ BAGAN$^1$, M.\ LAVELLE$^{1\,}$\footnote{Talk presented by 
M.\ Lavelle}, D.\ MCMULLAN$^2$, B.\ 
FIOL$^1$, N.\ ROY$^1$ }

\address{{}$^1$ IFAE, Edificio Cn, Universitat Autonoma de Barcelona,\\ 
E-08193 Bellaterra (Barcelona), Spain}

\address{{}$^2$ Department of Mathematics and Statistics, University of 
Plymouth, Drake Circus,\\ Plymouth, Devon PL4 8AA, UK}


\maketitle\abstracts{This talk reports on work aimed at improving our 
understanding of charged states in gauge theories. 
Emphasis is placed on different ways of implementing 
the gauge invariance of physical states. 
QED perturbative calculations are used to
stress that 
gauge invariance is not enough: only a subset of 
such states have a direct physical 
significance. In QCD a non-perturbative obstruction 
means that quarks and gluons cannot form physical states and as such 
these particles are not truely observables. This sets a fundamental 
limit on the constituent quark model.}
  
\section{Charged States in QED}

Gauss's law tells us that physical states in gauge theories have to be 
locally 
gauge invariant. This has many implications for any discussion of 
charged particles --- such as electrons or quarks\,\cite{monst}. 
In QED the basic fermion is not gauge invariant but one way of 
making it so is to attach a \lq string\rq\ to it:
\begin{equation}
\psi_{\Gamma}(x):=\exp\left(ie\int^x dz_\mu A^\mu(z) \right)\psi(x)\,, 
\label{eq:qedstring}
\end{equation}
where the line integral is over some path, $\Gamma$, from infinity to 
$x$.  
While this is 
gauge invariant if the gauge transformations fall off to constants at 
infinity, it has no physical signicance since the flux is compressed on 
to the path of the string. 
However, we may use 
it as a gauge invariant initial state and watch it evolve. If we 
assume that the charged particle is very heavy, i.e., static, then the 
Hamiltonian is just ${{1}\over{2}}\int E_i^2+B_i^2$, and the evolution 
is exactly solvable\,\cite{shab}. 
Essentially what happens is that the charge generates 
a Coulombic field and the string radiates away to infinity. 

To see how to describe this final state we recall\,\cite{dir} 
that in QED
\begin{equation}
\psi_{\rm f}(x):=\exp\left( ie\int d^4zf_\mu(z,x)A^\mu(z)\right) 
\psi(x)\,,
\label{eq:fcond}
\end{equation}
is gauge invariant if $\partial_\mu^z f^\mu(z,x)= 
\delta^{(4)}(z-x)$ holds. Clearly there are many possible solutions to 
Eq.~\ref{eq:fcond} and indeed Eq.~\ref{eq:qedstring} may be understood 
in this way. 

Another solution to Eq.~\ref{eq:fcond} is given by $f_0=0$, 
$f_i=4\pi\delta(z_0-x_0)\partial_i^z1/\vert{\boldmath z}-{\boldmath 
x}\vert$, with 
which we have  
\begin{equation}
\psi_{\rm c}(x):= 
\exp\left(ie{{\partial_iA_i}\over{\nabla^2}}(x)\right) \psi(x)\,.
\label{eq:psicdef}
\end{equation}
This is so defined as to be local in time, which has obvious 
advantages, but it is spatially non-local. It may be seen that this 
is in fact 
a {\em necessary} consequence of Gauss's law for any description of 
a physical charged state. (An operator describing a 
charged state cannot commute with the charge operator, but this last 
can with the help of Gauss's law be rewritten as: 
$\int\rho\to\int\partial_i E_i$ and this in turn may be re-expressed 
as a surface integral over the sphere at infinity. A description of a 
charged particle must therefore include a non-local electromagnetic 
cloud, a dressing, around the charge. The non-covariance of 
Eq.~\ref{eq:psicdef} is justified in detail elsewhere\,\cite{monst}.) 
Using the fundamental 
commutator, $[E_i({\boldmath x}), A_j({\boldmath 
y})]=i\delta_{ij}\delta^{(3)}({\boldmath 
x}-{\boldmath 
y})$, one finds that the commutator of the electric field with 
the charged field of Eq.~\ref{eq:psicdef} yields the field times the 
change in the electric field we would expect of a static charge. This 
is what Eq.~\ref{eq:qedstring} actually evolves into in the  
study described above. The extension of Eq.~\ref{eq:psicdef} to charges 
with a constant velocity,  
i.e., a gauge invariant description whose commutators with $E_i$ and 
$B_i$ yield the electric and magnetic fields associated with a 
charge moving with velocity ${\boldmath v}=(v^1,0,0)$
has been recently 
developed\,\cite{monst}:
\begin{eqnarray}
&&
\!\!\!\!\!\!\!\!\!\!\!\!
\psi_{{\boldmath v}}(x)= \exp\Bigg(\Bigg.
-\frac{e}{4\pi}\frac1{\sqrt{1-v^2}}\label{eq:moverdef}\\ 
&&\!\!\!\!\!\!\!\!\!\!\!\!
\times\int d^3z\frac{(1-v^2)\partial_1A_1(x^0,{\boldmath 
z})+\partial_2A_2(x^0,{\boldmath
z})+\partial_3A_3(x^0,{\boldmath z})-vE_1(x^0,{\boldmath
z})}{\vdenom{x}{z}^{\!\!\frac12}}\Bigg.\Bigg) \psi(x)\,. 
\nonumber 
\end{eqnarray}
It should be noted 
that this is not just a Lorentz boost of Eq.~\ref{eq:psicdef} and 
further that a gauge invariant term (dependent on $E_i$) is present 
here. This last point further 
exemplifies the fact that gauge invariance is only a 
necessary requirement for a physical state and that, e.g., energy 
considerations must also be taken into account in any more realistic 
description. 


To test the above interpretation of the dressed charges and to study 
the practicability of working with such variables various perturbative 
calculations have been carried 
out\,\cite{monst}$^{\!,\,}$\cite{slow}$^{\!,\,}$\cite{fast}. 
Recall that 
the mass-shell renormalisation of the usual fermion propagator in QED 
is, in general, marred by an infra-red divergence and that this 
divergence is understood as due to a neglect of the electromagnetic 
cloud surrounding the charge. The two-point function of the dressed 
charges was studied for the above-mentioned dressings which, it is 
again stressed, are supposed to correspond to charges moving with a 
particular velocity. This leads to the {\em prediction} that there will 
be no infra-red divergence if the propagator is taken on-shell at 
$p=m\gamma(1,v^1)$. This prediction has been 
{\em verified} by explicit calculation at one loop for an 
arbitrary velocity\cite{slow}$^{\!,\,}$\cite{fast}.  
Since the infra-red divergence is due to the 
slow fall-off of the Coulombic interaction, the anomalous magnetic 
moment of the electron does not contribute. This means that the same 
dressing should lead to infra-red finite results for both fermionic and 
scalar QED. This has also been verified explicitly. 

\section{Charged States in QCD}

A difference between QCD and QED is that the colour charge of the 
former theory 
\begin{equation}
Q^a=\int d^3{\boldmath x}\left( J^a_0({\boldmath 
x}) -f^{abc}E_i^b({\boldmath x}) 
A_i^c({\boldmath x}) \right)\,,
\label{eq:colchge}
\end{equation}
is not locally gauge invariant. It can, however, be 
demonstrated\,\cite{monst}$^{\!,\,}$\cite{colour} that it has a gauge 
invariant action on gauge invariant states. (One may rewrite the colour 
charge with the help of Gauss's law and then 
express it as a surface integral; 
if local gauge transformations fall off at spatial infinity to unity, 
the action of the charge is gauge invariant.) It 
follows\,\cite{monst}$^{\!,\,}$\cite{colour}  
that the colour statistics of coloured quarks 
dressed with coloured glue to form a gauge invariant whole will be 
just that of the naive quark model. So how do we dress the quarks? 
In perturbation theory 
one can extend the analysis of Eq.~\ref{eq:fcond} to QCD 
order by order. Another 
way to extract the 
equivalent of Eq.~\ref{eq:psicdef} is to consider a quark field with a 
path ordered exponential attached to it --- the QCD equivalent of 
Eq.~\ref{eq:qedstring}. It is possible order by order in perturbation 
theory to factor out the dependence on the path of the string, 
$\Gamma$. This yields at order $g^2$ for a path fixed in one time 
slice:  
\begin{equation}
\psi_\Gamma(x):={\cal P}\exp\left( g\int^x dz_iA_i(z) \right)\psi(x)
\to {\cal N}_\Gamma(x) \psi_{\rm c}^{g^2}(x) +O(g^3) \,, 
\label{eq:qcdstring}
\end{equation}
where ${\cal N}_\Gamma(x)$ is path dependent but gauge invariant to 
this order in $g$ and 
\begin{eqnarray} 
\psi_{\rm c}^{g^2}(x)&\!\!\!:= &\!\!\! \Bigg(\Bigg.
1+g  {{\partial_i A_i}\over{\nabla^2}}(x) + {{g^2}\over{2}}
\left(  {{\partial_i A_i}\over{\nabla}}(x) 
\right)^2  \label{eq:psicqcd}\\ & &\!\!\!\!\!
\!\!\!\!\!\!\!\!\!\!\!\!\!\!\!\!\!\!\!
- {{g^2}\over{2}}f^{abc}T^a {{1}\over{\nabla^2}}\left( A_j^b 
 {{\partial_j\partial_iA_i^c}\over{\nabla^2}} \right)(x)  
- {{g^2}\over{2}}f^{abc} 
T^a {{\partial_j}\over{\nabla^2}}\left( A_j^b 
 {{\partial_iA_i^c}\over{\nabla^2}} \right)(x)  
\Bigg.\Bigg)\psi(x) \,.
\nonumber
\end{eqnarray}
This is to order $g^2$ 
a gauge invariant, path independent description of a quark and 
the natural extension of Eq.~\ref{eq:psicdef}. 

It is useful to recall that in QCD we have, beyond lowest order in 
perturbation theory, no equivalent of the Coulombic electric field. 
This means that after finding gauge invariant dressings, it is not easy 
to understand what physical significance they have. Perturbative 
studies, energy minimisation etc.\ will all have their role to play. 
However, this talk will conclude by pointing out a fundamental 
obstruction to the construction of any gauge invariant dressed quark. 
In general a gauge invariant dressed quark may be written as
\begin{equation}
\psi_{\rm h}(x)= h^{-1}(x)\psi(x)\,,\:{\rm if\; under\; a\; 
gauge\; 
transformation\;} h\to h^U=U h\,.
\label{eq:hdef}
\end{equation}
This same field dependent 
$h$ also defines a gauge invariant gluon. The question is now 
how do we find $h$? It may be shown\,\cite{monst} that the 
above transformation property 
of such an $h$ means that not only may a gauge choice be used to 
construct an $h$, but any $h$ may be used to construct a gauge fixing. 
We see now that Eq.~\ref{eq:psicqcd} corresponds to Coulomb gauge. 
Since we know that good gauge fixings in QCD do not exist (the 
Gribov ambiguity) it is impossible to construct a physical quark state. 
This naturally explains why isolated quarks are not observed: such 
states cannot be constructed outside of perturbation theory. For 
more 
details and some suggestions as to how to obtain the scale at which the 
constituent quark model breaks down, we refer to Ref.\ 1.

\section*{Acknowledgments}
EB \& BF were supported by 
CICYT research project AEN95-0815 and  
ML by AEN95-0882. NR was supported 
by a grant from the region Rh\^one-Alpes.


\section*{References}
\begin{thebibliography}{99}
\bibitem{monst}For a review see: M.\ Lavelle and D. McMullan, {\em 
Constituent Quarks from 
QCD}, to appear in Physics Reports C.

\bibitem{shab}L.V. Prokhorov, D.V. Fursaev and S.V. Shabanov, 
\Journal{TMP}{97}{1355}{1994}

\bibitem{dir}P.A.M. Dirac, \Journal{\CJP}{33}{650}{1955}

\bibitem{slow}E. Bagan, M. Lavelle and D. McMullan, 
\Journal{\PLB}{370}{128}{1996}.

\bibitem{fast}E. Bagan, M. Lavelle and D. McMullan, {\em A Class of 
Physically Motivated Gauges with an Infra-Red Finite Electron 
Propagator}, submitted 
for publication. (hep-th/9602083). 

\bibitem{colour}M. Lavelle and D. McMullan, 
\Journal{\PLB}{371}{83}{1996}.

\end{thebibliography}
\end{document}